\documentclass[aps,prb,reprint,superscriptaddress]{revtex4-2}
\usepackage{amsmath}
\usepackage{amssymb}
\usepackage{graphicx}
\usepackage[colorlinks=true,linkcolor=blue,urlcolor=blue,citecolor=blue]{hyperref}
\date{\today}
\begin{document}
\title{Emergence of topological defects and spin liquid in a two-orbital spin-fermion model on the honeycomb lattice}
\author{Kaidi Xu}
\author{Shan-Shan Wang}
\email{wangss@seu.edu.cn}
\affiliation{Key Laboratory of Quantum Materials and Devices of Ministry of Education, School of Physics, Southeast University, Nanjing 211189, China}
\author{Rong Yu}
\email{rong.yu@ruc.edu.cn}
\affiliation{Department of Physics and Beijing Key Laboratory of Opto-electronic Functional Materials \& Micro-nano Devices, Renmin University of China, Beijing 100872, China}
\author{Shuai Dong}
\email{sdong@seu.edu.cn}
\affiliation{Key Laboratory of Quantum Materials and Devices of Ministry of Education, School of Physics, Southeast University, Nanjing 211189, China}

\begin{abstract}
Stabilizing exotic quantum phases of matter, e.g. spin liquid, is an attractive topic in condensed matter.  Here, by a Monte Carlo study of a two-orbital spin-fermion model on a honeycomb lattice, we show the cooperative effects of the orbital degeneracy of itinerant electrons and the exchange interaction of localized spins can significantly suppress both ferromagnetic and antiferromagnetic orders by generating topological defects and give rise to an intermediate spin liquid state via continuous phase transitions. This phase competition can also be achieved by tuning the electron filling. These results shed new light on realizing spin liquids on geometrically non-frustrated lattices.
\end{abstract}
\maketitle

\section{Introduction}
The search for exotic magnetically disordered ground states, including both 
classical and quantum spin liquids (SL's)~\cite{anderson_resonating_1973, doi:10.1126/science.aay0668, balentsspin2010, Savary_2017, Udagawa_Jaubert_2021}, is a long-going theme in the condensed matter community, and tremendous progress has been made in frustrated spin systems~\cite{Diep_book, 
lacroix2011introduction}. However, besides several well known examples, such as the Kitaev model on the 
honeycomb lattice~\cite{KITAEV20062}, SL states in microscopic spin models relevant to real materials are yet to be justified. Meanwhile, this motivates another route in searching for SL's -- by examining strongly correlated systems with itinerant electrons, for example, in Hubbard models in proximity to the Mott transition, or periodic Anderson models in certain limits~\cite{PhysRevB.71.144508, PhysRevB.78.045109, PhysRevLett.102.186401, PhysRevB.81.245121, PhysRevLett.95.036403, PhysRevB.83.235119}. Along this line, efforts have been made by considering systems with multiple orbitals and higher $\mathrm{SU}(N)$ ($N>2$) symmetries~\cite{PhysRevA.104.043316, PhysRevB.104.115119, PhysRevB.106.205129, PhysRevB.58.R4199, PhysRevLett.115.136401, PhysRevLett.124.016401, stadler_hundness_2019}. 

Appealing though, including itinerant electrons in the model poses great challenges for unbiased numerical studies: Quantum Monte Carlo (MC) simulations would suffer severe sign problem \cite{PhysRevLett.94.170201} when fermions of the system are away from half-filling and density matrix renormalization group and/or tensor network methods are limited by the rapid increase of entanglement entropy of systems in spatial dimensions higher than one~\cite{RevModPhys.93.045003}. Consequently, stabilization of SL's in these systems remain under debate~\cite{Meng2010, PhysRevLett.109.026404, PhysRevB.89.165134, Sorella2012, 
PhysRevLett.110.096402, PhysRevX.6.011029, PhysRevB.102.235105, PhysRevB.90.195133, PhysRevB.91.165108, PhysRevX.3.031010}, which motivate us to further explore for the key ingredients to realizing these highly nontrivial states of matter in models with coupling between localized magnetic moments and itinerant electrons.

\begin{figure}
\centering
\includegraphics[width=0.48\textwidth]{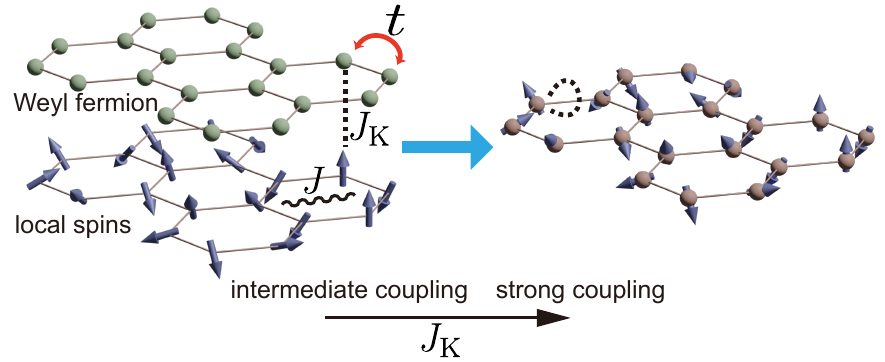}
\caption{Sketech of the two-orbital spin-fermion model on a bilayer honeycomb lattice. Itinerant electrons described by a two-orbital Weyl semi-metal model reside on the top layer of the lattice, with the hopping amplitude $t$. They are coupled to local spins (the bottom layer) via a ferromagnetic Kondo coupling $J_{\rm K}$. In the strong coupling limit ($|J_{\rm K}|\gg|t|$), spins of itinerant electrons are aligned by the local spins and their hopping integrals are renormalized by an emergent Berry phase $\Omega_{ij}$ at each bond. The neighboring spins are coupled with antiferromagnetic $J$.}
\label{fig:sch}
\end{figure}

The Kondo lattice model (KLM) provides the simplest picture for describing the interplay between itinerant 
electrons and local magnetic moments~\cite{10.1143/PTP.32.37, PhysRev.82.403, PhysRev.100.675, PhysRev.118.141, PhysRev.100.564, doi:10.1073/pnas.1715851115}. In recent years, SL phases in KLM have 
been proposed, which have attracted lot of attention~\cite{P_Coleman_1989, PhysRevB.66.045111, 
PhysRevLett.96.036601, PhysRevLett.98.026402, PhysRevB.106.115135, PhysRevB.99.035155, PhysRevB.107.L121111}. The competition between Kondo screening and magnetic correlations mediated by the 
Ruderman-Kittel-Kasuya-Yosida (RKKY) interactions serves as an additional source of quantum fluctuations that may cause deconfinement of spinons, novel quantum criticality, and non-Fermi liquid behavior~\cite{PhysRevB.39.7223, PhysRevB.44.2664, PhysRevLett.90.216403, PhysRevLett.71.1613, PhysRevB.97.085118, PhysRevLett.113.076401, PhysRevLett.74.4507, PhysRevLett.129.077202}. With including the orbital degree of freedom,  there are several well-established results that suggest orbital degeneracy may play the role similar to spin frustration and give rise to possible fractionalized Fermi liquids~\cite{PhysRevLett.90.216403} and SL-like behavior~\cite{PhysRevB.107.L121111}. Even when treating the local magnetic moments classically, highly nontrivial magnetic states can emerge, including spiral antiferromagnet,  skyrmion lattice (SkX)~\cite{Wang_PRL_2020}, as well as disordered cooperative paramagnet~\cite{PhysRevLett.104.106407}.  
  
In this work, we consider a reduced form of the Kondo-Heisenberg model by taking the local moments as 
classical spins, which is often termed as the spin-fermion model~\cite{PhysRevLett.84.2690, 
PhysRevLett.109.047001, doi:10.1126/science.283.5410.2034, PhysRevLett.80.845}. By studying a two-orbital spin-fermion model on a honeycomb lattice (see Fig.~\ref{fig:sch}), we show the interplay between orbital degeneracy of itinerant electrons and the antiferromagnetic superexchange between neighboring spins 
can significantly suppress long-range magnetic orders by inducing various topological defects (TDs) and drive the system into a SL state via continuous phase transitions. Remarkably, the SL state is stabilized within a certain regime of the phase diagram with tuning by either exchange coupling or electron filling. Our results suggest a new route in realizing SLs in correlated multiorbital systems on geometrically non-frustrated lattices.   

\section{Model \& method}
The Hamiltonian of the spin-fermion model 
we consider in this work reads as
\begin{equation}
\begin{split}
H &= \sum^{o,o^\prime=\{a,b\}}_{<ij>\sigma} t^{oo^\prime}_{ij} 
c^{\dagger}_{io\sigma}c_{jo^\prime\sigma} + J_{\rm K}\sum^{o}_{i\alpha \beta} 
c^{\dagger}_{io\alpha} \sigma_{\alpha \beta} c_{io\beta} \cdot 
\mathbf{S}_{i} \\
&+ J\sum_{<ij>} \mathbf{S}_{i} \cdot \mathbf{S}_{j} + A_{z} \sum_{i} 
(S^{z}_{i})^2,
\end{split}
\label{Eq:Ham}
\end{equation}
where $c^{\dagger}_{io\sigma}$ creates an itinerant electron in orbital $o$ with spin $\sigma$ at site $i$ of a honeycomb lattice. We consider a two-orbital model whose band structure describes a Weyl semimetal, and values of the nearest-neighboring hopping parameters $t^{oo^\prime}_{ij}$ of this model can be found in Ref.~\cite{PhysRevB.108.094401}. $\mathbf{S}_{i}$ is the local spin operator, which is treated here as a classical $O(3)$ vector. $J$ refers to the Heisenberg antiferromagnetic superexchange, and $A_{z}<0$ refers to an easy-axis single-ion 
anisotropy. $J_{\rm K}<0$ is a ferromagnetic Kondo coupling. As a generally accepted approxiamtion, $|J_{\rm K}| \gg |t|$, then the spin of itinerant electron is enforced to be in parallel to the local spin at that site. This leads to renormalization of the hopping integral $\tilde{t}^{oo'}_{ij} = \Omega_{ij} t^{oo'}_{ij}$ by a bond dependent Berry phase $\Omega_{ij}$, which is detailed in the Supplemental Material (SM)~\cite{sm}.

The Hamiltonian in Eq.~\eqref{Eq:Ham} can be solved  by combining MC and exact diagonalization techniques~\cite{PhysRevLett.80.845, PhysRevB.108.094401}. Within a fixed classical spin configuration $H (\{S_{i}\})$, it can be represented by a $2N \times 2N$ matrix where the fermion degree of freedom can be easily traced out by direct diagonalization by standard library routine. After this procedure, the effective action is left purely classical and then the spin configuration can be stochastically sampled by Metropolis algorithms. During every MC evaluation, the diagonalization is performed beforehand. In the numerical calculation, we take one hopping parameter $t^{bb2}=1$ as the energy scale. MC simulations are performed on lattices with linear dimension from $L=6$ to $L=16$ (total number of sites $N=2L^2$), with periodic boundary conditions. 
\begin{figure}
\centering
\includegraphics[width=0.48\textwidth]{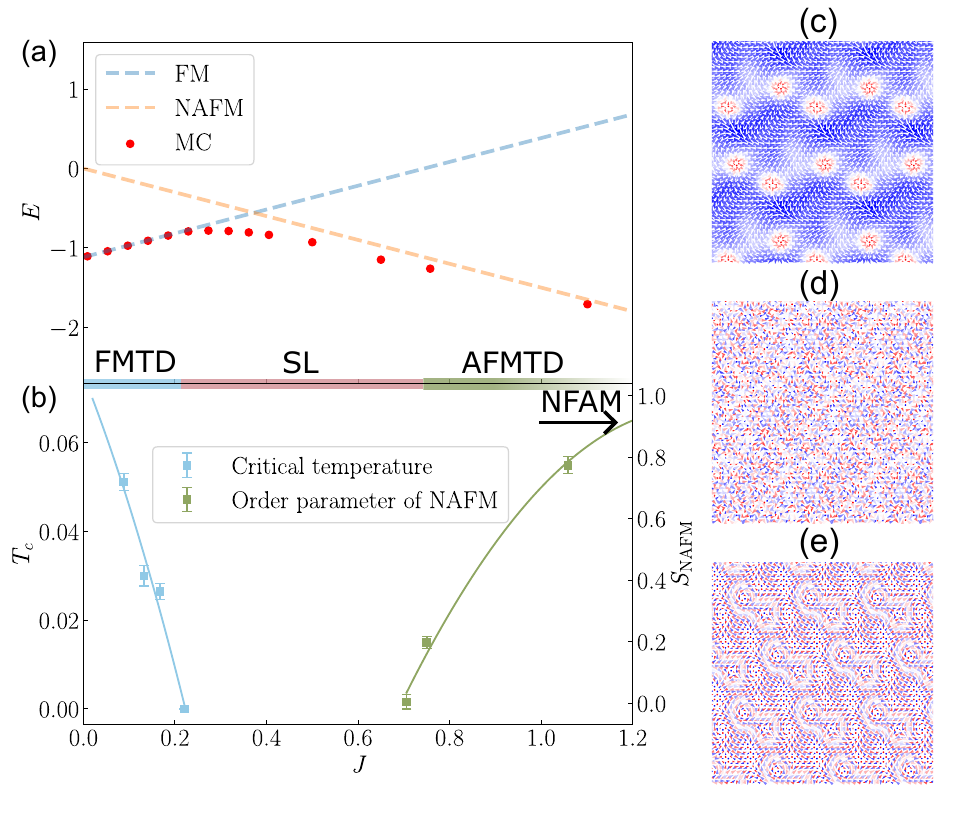}
\caption{(a) Evolution of the total energy $E$ as a function of superexchange coupling $J$ at quarter filling at (near) zero temperature. Dots: from MC simulations at extremely low temperature ($T=0.002$). Dashed lines: for perfect FM and NAFM orders at $T=0$. (b) The corresponding magnetic ground state phase diagram, determined from the evolution of FM $T_{\rm C}$ and the normalized static spin structure factor of the NAFM state. By increasing $J$, the ground state experiences a series of continuous phase transitions from the FM state first to a SL, then to a NAFM state. In either the FM or the NAFM phase, when close to the transition point to the SL phase, topological defects (TD's) of spin textures emerge which disturb the corresponding magnetic order. (c-e) Typical MC snapshots of spin textures at $T=0.002$, in the FM ($J=0.15$), SL ($J=0.50$), and NAFM ($J=1.12$) regions respectively. In (c) the TDs are skyrmion-antiskyrmion pairs, and in (e) they form antiferromagnetic skyrmion-antiskyrmion pairs.}
\label{fig:pd}
\end{figure}

\section{Results}
\subsection{Evolution of magnetic phases at quarter-filling}
First,  the evolution of magnetic ground states at quarter-filling is studied by varying the superexchange coupling $J$. As shown in Fig.~\ref{fig:pd}, for small $J$ the ground state has a FM order. As effects of the single-ion anisotropy, all spins are aligned along the $S^z$ direction. The Curie temperature $T_{\rm C}$ of FM phase decreases with increasing $J$ and becomes vanishingly small at $J\approx 0.22$. Above $J\approx 0.7$, the ground state develops a N\'{e}el AFM (NAFM) order, as demonstrated in Fig.~\ref{fig:pd}(b) by the the inflection point of the static spin structure factor $S(\mathbf{q})$ at the NAFM wave vector $\mathbf{q}=(\pi, \pi)$ \cite{sm}.

\begin{figure}
\centering
\includegraphics[width=0.48\textwidth]{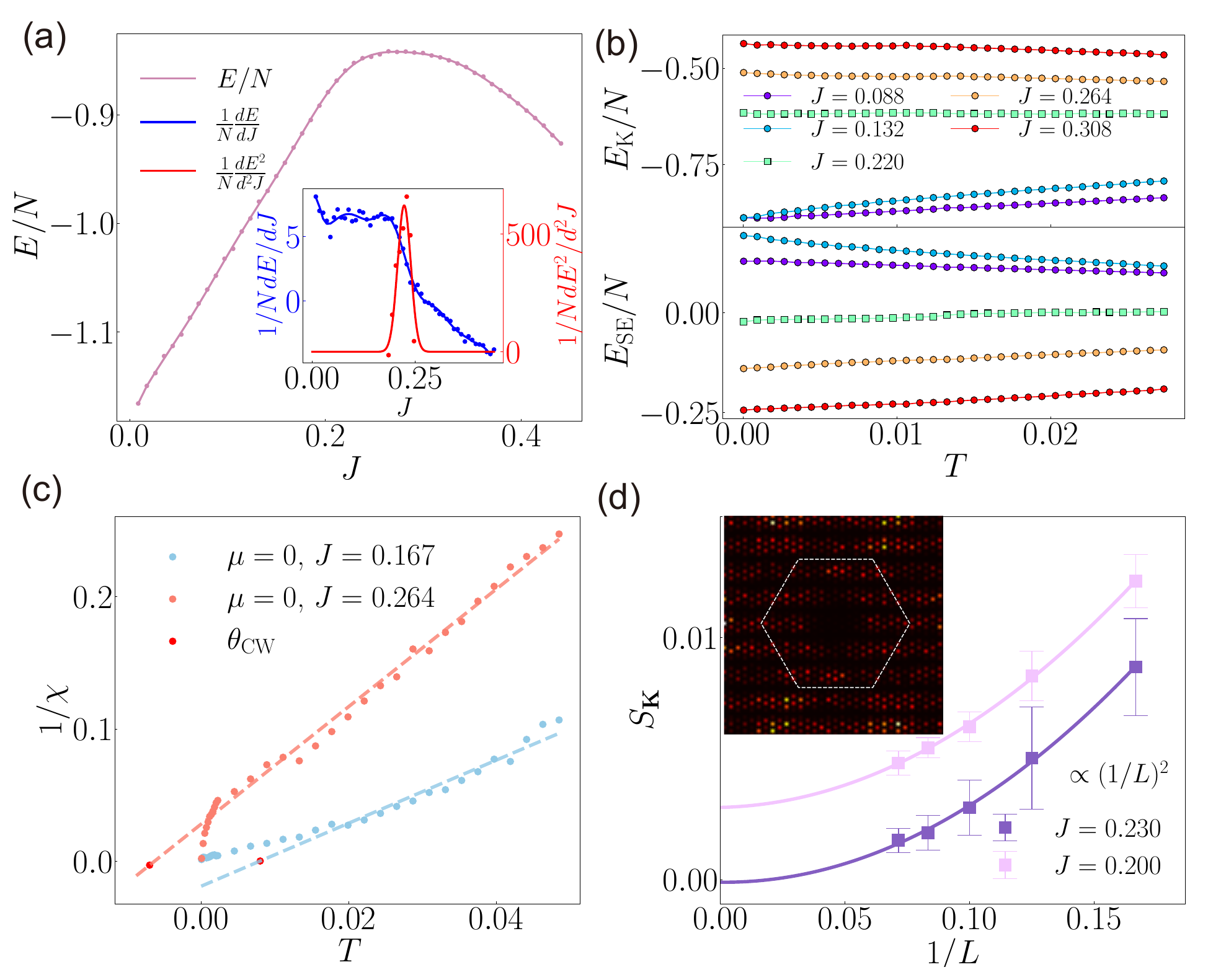}
\caption{(a) Evolution of the total energy $E$ with $J$ near the FM-to-SL transition. Inset: $dE/dJ$ and $d^2E/dJ^2$. Both $E$ and $dE/dJ$ vary continuously while $d^2E/dJ^2$ exhibits a peak at $J=0.22$, implying a second-order transition. (b) Temperature evolution of the kinetic energy $E_{\rm K}$ and the superexchange energy $E_{\rm SE}$ for selected $J$'s. (c) Inverse of spin susceptibility $1/\chi$ as a function of temperature for the FM and SL ground states. Lines are the Curie-Weiss fits to high-temperature data. (d) Finite-size scaling of the spin structure factor at $\mathbf{q}=(0, 0)$ for two $J$'s stabilizing the FM and SL ground states. Curves are quadratic fits. Inset: the static structure factor in the momentum space for $J=0.230$.}
\label{fig:2}
\end{figure}

Second, for $0.22\lesssim J\lesssim0.7$, the spins do not show any feature in either the real-space pattern [Fig.~\ref{fig:pd}(d)] or the structure factor in the momentum space [inset of Fig.~\ref{fig:2}(d)]. Moreover, our MC simulations do not find any spin freezing at low temperatures [see Fig.~S2 in SM~\cite{sm}]. This, together with properties discussed below, suggests the magnetically disordered ground state in this intermediate $J$ regime is a classical SL (sometimes also called a cooperative paramagnet \cite{PhysRevLett.80.2929}). 

As shown in Fig.~\ref{fig:2}(a), in the zero-temperature limit, the total energy $E$ and its first derivative $dE/dJ$ vary continuously across the FM-to-SL transition, but the second derivative $d^2 E/dJ^2$ exhibits a prominent peak. This implies that the FM-to-SL transition is second-order. Similar conclusion applies to the SL-to-NAFM transition. Note that the nature of these transitions is in sharp contrast to that in one-orbital spin-fermion models on geometrically non-frustrated lattices, in which tuning the superexchange coupling $J$ usually induces a strong first-order transition between a FM metal and an insulating NAFM state~\cite{PhysRevLett.80.845, PhysRevB.58.6403, PhysRevB.58.6414}. It is rather similar to the (A)FM-to-SL transition in frustrated lattice~\cite{PhysRevLett.104.106407, PhysRevLett.110.097204}, although our system is non-frustrated.

\subsection{Topological defects \& properties of SL}
To better understand these magnetic transitions in our model, we examine the MC snapshots in FM and NAFM states. Remarkably, we find the real-space spin patterns close to the transitions contain various TD's. In the FM phase, depending on the value of $J$, they can be either skyrmion-antiskyrmion pairs, meron-antimeron pairs, vortex-antivortex pairs, or even their combinations, as shown in Fig.~\ref{fig:pd}(c) and Fig. S2 in SM~\cite{sm}. In the NAFM phase, TD's also appear in pairs, as illustrated in Fig.~\ref{fig:pd}(e). The only difference is that spins on the two sublattices form interpenetrating skyrmions (or other TD's) with opposite chirality (coined as AFM skyrmion here). In both the FM and NAFM phases, TD's always appear in pairs, and each pair is formed by bounding one particle (skyrmion, meron, etc.) and its anti-particle counterpart with opposite chirality.

Recently, magnetic phases with nontrivial topological structures, such as skyrmion crystal (SkX) have been extensively studied. A well-known mechanism to stabilize these phases lies in the interplay among Dzyaloshinskii-Moriya interaction, geometric/exchange frustration, and magnetic anisotropy~\cite{Fert2017}. It is also noticed that in centrosymmetric KLM, Fermi surface nesting and out-of-plane external field also 
play crucial roles~\cite{PhysRevLett.118.147205, PhysRevLett.129.017201}. The TD's emerging in our model are different from these known phases. Although periodic structure of TD's can not be fully excluded due to the small lattice used in simulation, the pairwise appearance of the TD's and the monotonically reduced but nonzero $T_{\rm C}$ imply that they are topological excitations disturbing the magnetic orders. The emergence of TD's provides a likely scenario for the FM(NAFM)-to-SL transition, i.e. the SL phase arises from proliferation of TD's. Note here the easy-axis magnetic anisotropy excludes a Kosterlitz-Thouless (KT) transition. It would be then interesting to examine the universality of this transition. 

It should be noted that possible influence from finite-size lattice is checked via scaling (Fig.~\ref{fig:2}(d)). Besides, using a similar model \cite{PhysRevB.104.174432}, Matsui {\it et al.} proved that the size of topological defects is depends on the lifetime of itinerant electrons and coupling constant between electrons and localized spins, which are fixed in our simulation. Indeed, we did not find any evidence the topological defects is unstable against the lattice size. Furthermore, by employing a one-orbital model which allows much larger lattices, the topological defects remain robust (Fig. S4 in SM) \cite{sm}.

We then investigate the properties of SL phase. Figure~\ref{fig:2}(b) shows the temperature dependence of the kinetic and superexchange energies ($E_{\rm K}$ and $E_{\rm SE}$) per site at several $J$ values. For the FM phase, both energies show clear temperature dependence. Near the critical point $J=0.22$, both energies show minimal temperature dependence. Further increasing $J$ where the ground state becomes a SL, the temperature dependence of both energies remains weak. Note that the entropy is evaluated by the integration of $1/TdE/dT$, the weak temperature dependence of energies suggests a large residual entropy of the SL. Also note that compared to $E_{\rm SE}$, $E_{\rm K}$ shows even weaker temperature dependence in the SL phase. This implies that the residual entropy is mainly ascribed to the orbital fluctuations of the model, given that the lattice has no geometric frustration. This is in contrast to the case with geometric frustration~\cite{PhysRevLett.104.106407} where both $E_{\rm K}$ and $E_{\rm SE}$ show minimum temperature dependence only within a narrow critical regime.

We now turn to study the temperature dependence of the magnetic susceptibility $\chi$, which usually follows a Curie-Weiss behavior at high temperatures as $\chi=C/(T-\theta_{\rm CW})$, where $C$ is a constant and $\theta_{\rm CW}$ is the Curie-Weiss temperature. The Curie-Weiss fit provides not only an estimate of the exchange interaction but also a measurement of frustration. Figure~\ref{fig:2}(c) shows the $1/\chi$ 
for two $J$ values below and above the FM-to-SL transition, respectively. For $J=0.167$ (in the FM region), the fitted Curie-Weiss temperature $\theta_{\rm CW}\approx0.01$, indicating an effective FM interaction among spins. Moreover,  $1/\chi$ deviates from the Curie-Weiss behavior at $T\approx0.02$. This temperature is slightly higher than $\theta_{\rm CW}$, indicating a rather weak frustration effect for this $J$. In contrast, for $J=0.264$ (in the SL region), the fitted $\theta_{\rm CW}\approx-0.01$, showing that the effective exchange interaction changes to AFM. Interestingly, for $T\ll |\theta_{\rm CW}|$, $\chi$ undergoes a Curie crossover to $1/\chi\sim T$ behavior. Such a crossover behavior indicates strong spin frustration and has been observed in a number of SL models~\cite{PhysRevB.108.024411} and SL candidate materials~\cite{PhysRevLett.98.107204}. In addition, the finite-size scaling analysis in Fig.~\ref{fig:2}(d) indicates that within this SL regime, the magnetic order parameter drops to zero in the thermodynamic limit, confirming that the system is magnetically disordered. 

\begin{figure}
\centering
\includegraphics[width=0.48\textwidth]{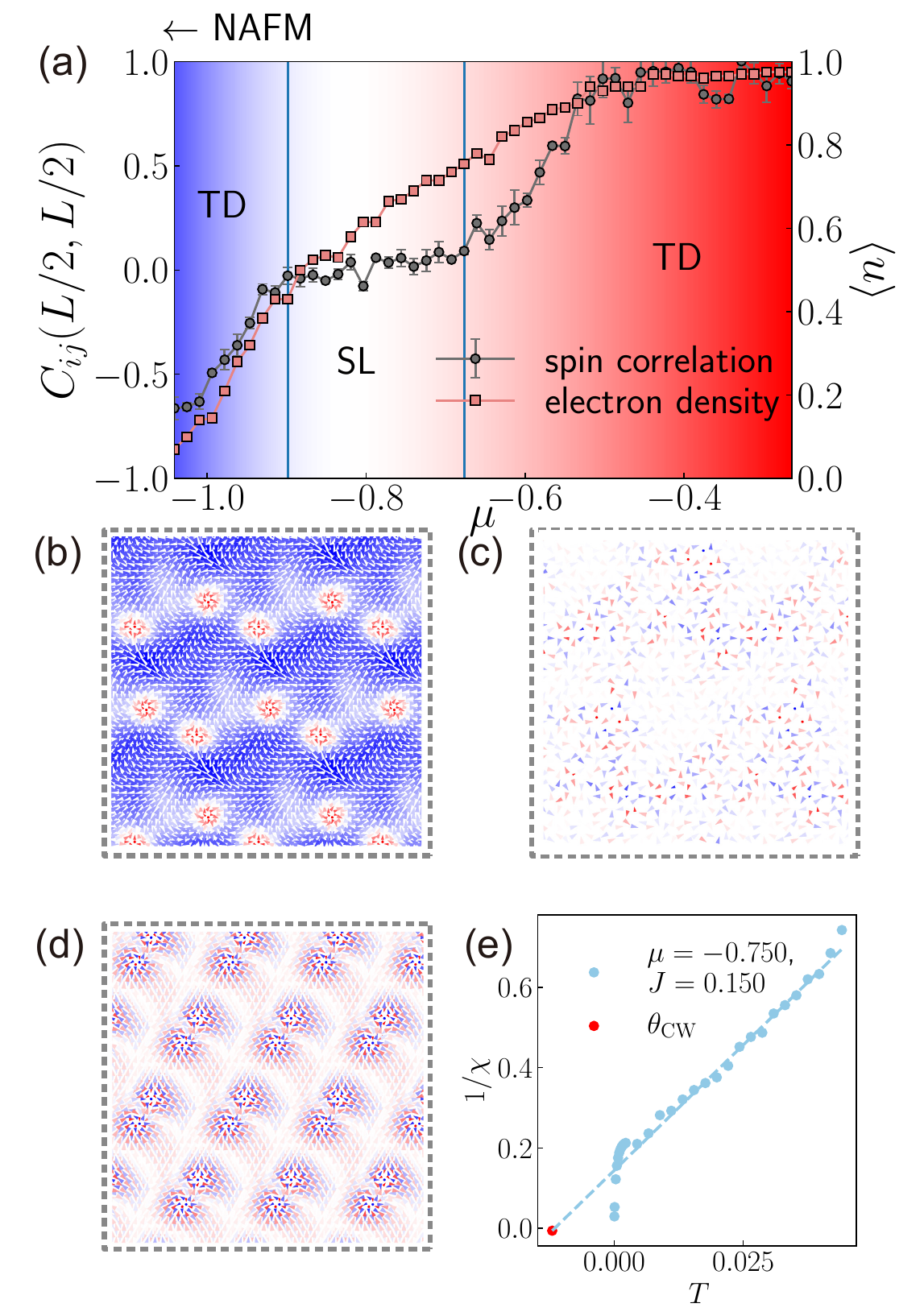}
\caption{The phase diagram as a function of chemical potential with fixed $J=0.15$ and at $T=0.002$. The squares represent the density of itinerant electrons per site while the circles represent the real-space correlation function. (b-d) The MC snapshots for $\mu=-0.95$, $-0.8$, and $-0.5$, respectively. (e) The inverse susceptibility for $\mu=-0.75$.}
\label{fig:pd-s}
\end{figure}

\section{Discussion}
Besides the superexchange coupling $J$, the FM-SL-AFM transition can also be induced by tuning the chemical potential $\mu$. Figure~\ref{fig:pd-s} shows a typical phase diagram of the system with $\mu$ for $J=0.15$. To trace the variation of magnetic order with $\mu$, we calculate the spin correlation function $C(\mathbf{R})=\langle \mathbf{S}_{i}\cdot \mathbf{S}_{i+\mathbf{R}}\rangle$ at the largest spatial distance $\mathbf{R}=(L/2, L/2)$. At the quarter-filling (i.e. electron density per site $\langle n \rangle=1$), the ground state is FM  (i.e. $C(\mathbf{R}) \approx 1$). With decreasing $\mu$, the electron density decreases continuously, and TD's appears [see Fig.~\ref{fig:pd-s}(b)]. For $\langle n \rangle\lesssim 0.8$, $C(\mathbf{R})$ develops a zero plateau, indicating the suppression of the FM order and the emergence of the SL. This SL phase is stabilized till $\langle n \rangle \approx 0.5$ ($1/8$-filling). In analogy to the quarter-filling case, this SL exhibits a disordered spin 
pattern [Fig.~\ref{fig:pd-s}(c)] and a Curie crossover in the susceptibility [Fig.~\ref{fig:pd-s}(e)]. 

Further decreasing the chemical potential yields the NAFM phase, as characterized by a negative $C(\mathbf{R})$. Interestingly, AFM TD's are 
observed in the NAFM phase, as shown in Fig.~\ref{fig:pd-s}(d). Stabilization of the SL within a finite range of electron filling suggests that this phase is insensitive to the shape of the Fermi surface, which is known to be relevant for weak Kondo couplings~\cite{Wang_PRL_2020}. In our case, the SL phase instead emerges as a compromise of the competing FM double-exchange and AFM superexchange interactions. In the vicinity of the SL phase, TD's appear as a consequence of strong fluctuations induced by this competition, which are further enhanced by the nontrivial topology of the bandstructure. 

As a final remark, it has been known that there doesn't exist an intermediate regime for SL phase in the one-orbital Dirac fermion model even if the correlation effect is considered \cite{PhysRevX.6.011029, PhysRevX.10.031016}, namely the electron correlation alone is not enough in searching for such highly-entangled state of matter. Then it is of great interest to examine whether the degenerate orbitals can introduce another flavor of gauge field fluctuation, a defining characteristic of a SL. Here, a two-orbital basis on the honeycomb lattice has been tested, which naturally manifests a Weyl-type band structure in the reciprocal space. In the spin-fermion model (one-orbital or two-orbital), the exchange frustration from antiferromagnetic $J$ can naturally tune the ground state from ferromagnetism to antiferromagnetism, with topological defects in real space emerging during this transition. The most interesting physics is that only in the two-orbital version the SL phase can survive as an intermediate phase, in which the orbital-degeneracy plays the essential role. This scenario is summarized in Fig. S3 in SM, and the controlled study of the one-orbital model can be found in Fig. S4 in SM \cite{sm}.

\section{Conclusion}
To summarize, in our model, topological defects have been observed in both ferromagnetic and antiferromagnetic phases in a two-orbital spin-fermion model on the honeycomb lattice. These topological defects suppress magnetic orders and eventually lead to an intermediate spin liquid phase via continuous quantum phase transitions, which can be achieved by tuning either the superexchange interaction or the electron density. Our results suggest that fluctuations introduced by orbital degeneracy can be a possible recipe for preparing a highly nontrivial quantum state on a geometrically nonfrustrated lattice. The model proposed here has the potential to be realized in the recently discovered itinerant magnetic materials on the hexagonal lattice with partially filled $e_{\rm g}$ manifold \cite{doi:10.1126/sciadv.aau3402}, dilute honeycomb magnets \cite{Lee2023}, or the ultracold atomic optical lattice which can mimic the SU($N$) ($N \geq 2$) double-exchange model \cite{Gorshkov2010}.

\begin{acknowledgments}
K. X. would like to thank Profs. Sergio Ciuchi and Sankalp Kumar for enlightening discussions. This work was supported by the Natural Science Foundation of China (Grants Nos. 12325401 \& 12104089 \& 12274069), and the Big Data Computing Center of Southeast University. The work at Renmin University is supported by the National Key R\&D Program of China (Grant No. 2023YFA1406500) and the National Natural Science Foundation of China (Grant Nos. 12334008 and 12174441).
\end{acknowledgments}
\bibliography{cite}
\end{document}